\documentclass[11pt]{article}
\usepackage{epsfig}
\usepackage{latexsym}

\def\bg#1{\mbox{\boldmath$#1$}}

\newcommand{\anabla}{{\overrightarrow{\nabla}}\!\!\!\!\!\!{\overleftarrow{\nabla}}}

\newcommand{\del}{\partial}
\newcommand{\beq}{\begin{eqnarray}}
\newcommand{\eeq}{\end{eqnarray}}
\newcommand{\be}{\begin{eqnarray*}}
\newcommand{\ee}{\end{eqnarray*}}
\newcommand{\bk}{{\bf k}}
\newcommand{\bp}{{\bf p}}
\newcommand{\bq}{{\bf q}}

\newcommand{\ra}{\rightarrow}
\newcommand{\e}{\epsilon}

\newcommand{\nn}{\nonumber}
\newcommand{\ket}[1]{\mbox{$\mid\!#1\rangle$}}
\newcommand{\bra}[1]{\mbox{$\langle#1\!\mid$}}

\newcommand{\ex}[1]{\langle\,#1\,\rangle}

\newcommand{\bpi}{{\bg\pi}}

\begin{document}

\centerline{\Large\bf {Non-Relativistic Pion Interactions and the Pionium Lifetime}}
\vskip 10mm
\centerline{Xinwei Kong}
\medskip
\centerline{\it Institute of Physics, University of Oslo, N-0316 Oslo, Norway.} 
\vskip 10mm
\centerline{Finn Ravndal\footnote{Permanent address: Institute 
            of Physics, University of Oslo, N-0316 Oslo, Norway}} 
\medskip
\centerline{\it NORDITA, Blegdamsvej 17, DK-2100, Copenhagen \O .}

\bigskip
\vskip 5mm
{\bf Abstract:} {\small We construct an effective Lagrangian for interacting pions
with non-relativistic energies. The coupling constants can be expressed in terms of the
different scattering lengths and slopes. When used in the calculation of the pionium
decay rate, the scattering slope contribution gives a correction of about 8\% compared
with the lowest order contribution coming from the scattering lengths alone.}


\bigskip

Pionium is a hadronic atom of $\pi^+$ and $\pi^-$ bound by the Coulomb force. It is
highly unstable via the strong decay $\pi^+ + \pi^- \ra \pi^0 + \pi^0$ which probes
the low-energy interactions of the pions. As such it can be used to test more
accurately the predictions of chiral perturbation theory which is an effective theory
for QCD at low energies\cite{Weinberg_1}\cite{Ecker}. It was first constructed by 
Weinberg who used it at tree-level to calculate the $\pi \pi$ scattering amplitudes 
in agreement with current algebra results\cite{Weinberg_2}. Since then the results
have been improved with one-loop corrections by Gasser and Leutwyler\cite{GL} and are
now carried to two-loop order\cite{2loop}. On the other hand, the experimental
values of these scattering amplitudes are still very uncertain. For instance, the 
isospin-zero $S$-wave scattering length is known with only 20\% accuracy\cite{DGH}.

Recently a lot of interest has been generated by the possibility of a more accurate
determination of scattering lengths from measurements of the hadronic 
decay of pionium\cite{many}\cite{Labelle}. In order for
this to succeed, one must have a complete understanding of the different effects
acting in the decay process. Since the pions in this hadronic atom are non-relativistic,
they can be described by an effective theory expanded in terms of operators of
increasing dimensions involving pion fields and their derivatives. By matching it to
relativistic chiral perturbation theory or experiments, the {\it a priori} unknown
coupling constants can be determined. It will be done in the following to order 
${p^2}$ where $p$ is the momentum of the pions. To this order the different
scattering amplitudes are characterized by a scattering length $a$ and a scattering
slope $b$ which then determine the coupling constants.

The same, effective Lagrangian also determines the dynamics of the bound pions in
the pionium atom. Since we have a strictly non-relativistic system, we can use the
ordinary Schr\"odinger equation to calculate the wavefunctions and there is no need
for covariant formalisms like the Bethe-Salpeter equation or others. This is in the 
very spirit of NRQED established by Caswell and Lepage\cite{CL}\cite{KL} and used 
with great success for muonium\cite{KN} and positronium\cite{HLZ}. In this 
non-relativistic framework one can then systematically calculate corrections to the
different energy levels. In particular, a complex contribution $\Delta E$ signals 
that the corresponding state is unstable with a decay rate given by $\Gamma = 
- 2\,\mbox{Im}\, \Delta E$ and thus with lifetime $\tau = 1/\Gamma$.

The dominant part of the pionium decay comes from the
constant part of the amplitude for $\pi^+ + \pi^- \ra \pi^0 + \pi^0$, i.e. from the
scattering length. In the following we will show that the energy dependence of the
amplitude, or the scattering slope, gives an additional contribution which is around
8\% of the leading term. It is an important correction and larger than typical
electromagnetic corrections which have been considered until now\cite{many}\cite{Labelle}.

Non-relativistic pions are described by the complex Schr\"odinger fields $\bpi = 
(\pi_+, \pi_0, \pi_-)$ where $\pi_+$ annihilates a $\pi^+$, $\pi_-^*$ creates a $\pi^-$
and so on. The free fields are described by the Lagrangian
\beq
    {\cal L}_0(\pi_i) = \pi_i^*\left(i{\del\over\del t} 
              + {1\over 2m_i}{\bg\nabla}^2\right)\pi_i
\eeq
The masses of $\pi_+$ and $\pi_-$ are the same and will be denoted by $m_+$ while
$\pi_0$ has a slightly lower mass denoted by $m_0$. We could also include here a 
relativistic coupling $\propto \pi_i^*{\bg\nabla}^4\pi_i$, but we will ignore such
small corrections in the following. In the same vain, we will not consider 
electromagnetic effects although they are in general important in the problem 
under consideration.

For  the interacting part ${\cal L}_{int}$ we will assume exact isospin invariance and
only $S$-wave interactions. 
We then find that the lowest order interaction can only involve two possible couplings,
\beq
    {\cal L}_{int}(\bpi) = G_0(\bpi^*\cdot\bpi)(\bpi^*\cdot\bpi) 
                           + H_0(\bpi^*\cdot\bpi^*)(\bpi\cdot\bpi)        \label{Lint}
\eeq
Thus, we have the full Lagrangian ${\cal L} = {\cal L}_0(\pi_+) + {\cal L}_0(\pi_0)
+ {\cal L}_0(\pi_-) + {\cal L}_{int}(\bpi)$. The interaction has dimension six and
is thus not renormalizable in the ordinary sense. But considered as an effective theory,
it can be renormalized to every order in the expansion of ${\cal L}_{int}$ in  
higher-dimensional operators. It has essentially the same form as a corresponding 
effective theory proposed for non-relativistic nucleons by
Weinberg\cite{Weinberg_3} and recently improved by Kaplan, Savage and Wise for $np$ 
scattering\cite{KSW_1} and the deuteron\cite{KSW_2}. The divergent loop integrals can 
be regularized by a momentum cutoff, but as for most effective theories, it is much more
efficient to use dimensional regularization with minimal subtraction. We
will use this method in the following.

For dimensional reasons we know that the coupling constants $G_0$ and $H_0$ must be 
$\propto 1/m^2$ where the 'heavy mass' $m$ in our case is the pion mass. They can be 
obtained by matching to relativistic chiral perturbation theory or directly to
experiments. Performing the matching in the first way, we find to lowest order in 
the expansion of the chiral Lagrangian the effective couplings
$G_0 = -1/8f_\pi^2$ and $H_0 = +3/16f_\pi^2$ with the pion decay constant $f_\pi = 
92.5$ MeV. The resulting couplings between pions in different isospin channels can
now be deduced from Eq.(\ref{Lint}) which takes the form
\beq
    {\cal L}_{int}(\bpi) &=& {1\over 4} A_0(\pi_0^* \pi_0^*\pi_0 \pi_0) 
                           + B_0(\pi_+^* \pi_-^*\pi_+ \pi_-) 
 + {1\over 2}C_0 (\pi_+^* \pi_-^*\pi_0 \pi_0 + \pi_0^* \pi_0^*\pi_+ \pi_-)  \nn \\
  &+& {1\over 4}D_0(\pi_+^* \pi_+^*\pi_+ \pi_+ + \pi_-^* \pi_-^*\pi_- \pi_-
   + 2\pi_+^* \pi_0^*\pi_+ \pi_0 + 2\pi_-^* \pi_0^*\pi_- \pi_0)       \label{L_int}
\eeq
when written out. With the above tree-level values for the two fundamental
coupling constants, we now have $A_0 = B_0/2 = C_0/3 = - D_0/2 = 1/4f_\pi^2$. 

In order to compare with experiments, we calculate the $S$-wave scattering 
amplitude $T(\bp)$ where 
$\bp$ is the CM-momentum of the pions. The real part is usually defined by
\beq
           \mbox{Re}\,T(\bp) = {8\pi\over m_\pi^2}
           \left(a + b{\bp^2\over m_\pi^2}\right)                      \label{ReT}
\eeq
in terms of the scattering length $a$ and the slope parameter $b$ which gives the 
energy dependence of the amplitude to lowest order\cite{DGH}. With the above values for the 
coupling constants, it is now straightforward to read off Weinberg's scattering 
lengths in the different isospin channels\cite{Weinberg_2} from the Lagrangian 
Eq.(\ref{L_int}).

Instead of using these results from chiral perturbation theory at tree level,
we can instead match the coupling constants to the measured cross sections, i.e. to
the observed $S$-wave scattering lengths $a_0$ and $a_2$ for isospin $I=0$ and $I=2$
respectively and the corresponding scattering slopes. 
\begin{figure}[htb]
 \begin{center}
  \epsfig{figure=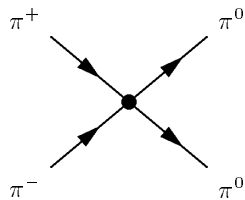,height=25mm}
 \end{center}
 \vspace{-8mm}            
 \caption{\small Tree-level contribution to the scattering process due to 
                 the contact term $C_0$.}
 \label{fig.1}
\end{figure}In connection with pionium we
will be especially interested in the process $\pi^+ + \pi^- \ra \pi^0 + \pi^0$. 
To lowest order in perturbation theory the scattering amplitude is given by the Feynman
diagram in Fig.1 which gives
\beq
     T^{(0)}(\pi^+ + \pi^- \ra \pi^0 + \pi^0) = {8\pi\over3 m_\pi^2}(a_0 - a_2)
     = C_0                                                             \label{C}
\eeq          
We thus have $C_0$ directly expressed in terms of measured scattering lengths. 
Considering related processes, we can similarly obtain the other coupling constants,
\beq
     A_0 = {8\pi\over3 m_\pi^2}(a_0 + 2a_2), \hspace{10mm}
     B_0 = {8\pi\over3 m_\pi^2}(a_0 + a_2/2), \hspace{10mm} 
     D_0 = {8\pi\over m_\pi^2} a_2                                  \label{ABD}
\eeq 
when we combine scattering amplitudes with definite isospin.
  
So far these relations are only valid at tree level of the effective theory. The
scattering amplitudes are real and unitarity is thus not satisfied. This can be
achieved by going to higher orders in perturbation theory.
\begin{figure}[htb]
 \begin{center}
  \epsfig{figure=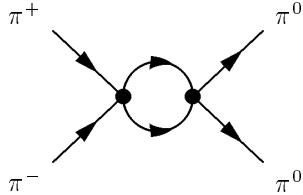,height=25mm}
 \end{center}
 \vspace{-8mm}            
 \caption{\small One-loop contributions to the scattering amplitude. }
 \label{fig.2}
\end{figure}
Again considering
$\pi^+ + \pi^- \ra \pi^0 + \pi^0$, we have two one-loop diagrams of the form shown
in Fig.2. One contains $\pi^+\pi^-$ and the other $\pi^0\pi^0$ in the intermediate 
state. Ignoring here the mass differences, they give the correction
\beq
    T^{(1)}(\pi^+ + \pi^- \ra \pi^0 + \pi^0)= -\left(B_0 C_0 + {1\over 2}A_0 C_0\right)
    I(\bp)                                                  \label{T1}
\eeq
where the factor of $1/2$ is due to the two identical particles in the $\pi^0\pi^0$ 
intermediate state. 

The integral over intermediate momenta
\beq
     I(\bp) = \int\!{d^3k\over (2\pi)^3}{1\over E - \bk^2/m_\pi + i\e}    \label{Ip}   
\eeq
where $E = \bp^2/m_\pi$ is the total CM energy, is seen to be linearly divergent. 
Using now dimensional 
regularization with minimal subtraction\cite{KSW_1}, it is simply given by 
$I(\bp) = - i|\bp| m_\pi/4\pi$. Using instead a momentum cutoff $\Lambda$, 
it would contain
a term proportional with $\Lambda$. This could then be absorbed by renormalization
of the coupling constant $C_0$. The net result is either way a purely imaginary result
which arises from the unitarity requirement, but does not contribute to the scattering
length or slope parameter in Eq.(\ref{ReT}).

However, going to two loops as in Fig.3 we will obtain corrections to the tree-level 
results. The two intermediate bubbles can again contain a $\pi^+\pi^-$ or a $\pi^0\pi^0$
pair. Summing up the four contributions from combinations of different bubbles, 
we then have the next order correction
\beq
    T^{(2)}(\pi^+ + \pi^- \ra \pi^0 + \pi^0)   &=& \left(
   {1\over 2}C_0^3 + {1\over 2}A_0B_0C_0 + {1\over 4}A_0^2C_0 + B_0^2C_0\right)
    I^2(\bp)                                                        \nn \\
   &=& - {8\pi\over 3m_\pi^4}(a_0 - a_2)(a_0^2 + a_0a_2 + a_2^2)p^2     \label{T2}  
\eeq
after regularization. 
\begin{figure}[htb]
 \begin{center}
  \epsfig{figure=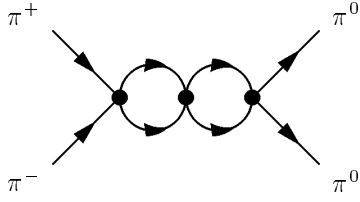,height=25mm}
 \end{center}
 \vspace{-8mm}            
 \caption{\small Two-loop contributions to the scattering amplitude.}
 \label{fig.3}
\end{figure}
It is seen to give an energy-dependence of the scattering
amplitude proportional to $p^2$ and thus contribute to the slope parameter in 
Eq.(\ref{ReT}). But such an energy dependence at two-loop level can also result
at tree level from an operator in the Lagrangian containing two derivatives. More
specifically, for the $\pi^+ + \pi^- \ra \pi^0 + \pi^0$ channel we consider here, 
we must include higher dimensional operators in the expansion of the effective 
Lagrangian. For $S$-wave interactions there is only one such possible operator to lowest
order in the derivative expansion,
\beq
     {\cal L}_{int}(\pi^+\pi^-\pi^0\pi^0) = {1\over 2}C_0 (\pi_+^* \pi_-^*\pi_0 \pi_0)
     + {1\over 4}C_2 (\pi_+^* \pi_-^*\pi_0 \anabla^2 \pi_0 + 
       \pi_+^* \anabla^2 \pi_-^*\pi_0 \pi_0) + \mbox{h.c.} \label{eft}
\eeq
Here the gradient is defined as $\anabla = 1/2(\overrightarrow{\nabla} - 
\overleftarrow{\nabla})$. It corresponds to
the vertex $\hat{V}_2$ in Fig.4 with the value $\bra{\bp}\hat{V}_2\ket{\bq} =  
{1\over 2}C_2(\bp^2 + \bq^2)$ in the CM reference frame. 
\begin{figure}[htb]
 \begin{center}
  \epsfig{figure=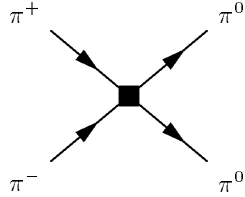,height=25mm}
 \end{center}
 \vspace{-8mm}            
 \caption{\small Vertex due to the double-derivative coupling $C_2$.}
 \label{fig.4}
\end{figure}
Adding this contribution to the two-loop result Eq.(\ref{T2}) and matching to the 
definition of the full scattering amplitude in Eq.(\ref{ReT}), we have
\be
   T^{(2)} - C_2p^2 = {8\pi\over 3m_\pi^2}{p^2\over m_\pi^2} (b_0 - b_2)
\ee
which gives
\beq
     C_2 = - {8\pi\over 3m_\pi^4}\left[(b_0 - b_2)  
           +(a_0 - a_2)(a_0^2 + a_0a_2 + a_2^2)\right]              \label{C2}
\eeq
Since this is a coupling constant in the effective Lagrangian, it can also be used 
for bound state problems. There are no 'off-shell' problems in this approach.

We are now in the position to consider decay of pionium. The ground state with 
wavefunction $\Psi(r)$ has the energy $E = 2m_+(1 - \alpha^2/8)$. It will be
perturbed by the different hadronic contact interactions in Eq.(\ref{L_int}). For 
instance, at tree level we get a real energy shift from the elastic coupling $B_0$.
Its magnitude is simply $-B_0|\Psi(0)|^2$ where the wave function at the origin is
$|\Psi(0)|^2 = \gamma^3/\pi$ with $\gamma = \alpha m_+/2$. 
It is proportional to the scattering length $a_0 + a_2/2$. This is the 
hadronic energy level shift discussed first by Deser, Goldberger, Baumann and 
Thirring\cite{Deser}. 

At next order in perturbation theory
we must evaluate the diagram in Fig.5 with a $\pi^+\pi^-$ in the intermediate state.
It gives also a real, but smaller contribution proportional to $(a_0 + a_2/2)^2$. 
\begin{figure}[htb]
 \begin{center}
  \epsfig{figure=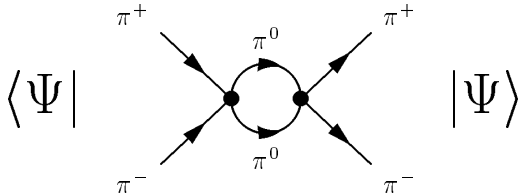,height=25mm}
 \end{center}
 \vspace{-8mm}            
 \caption{\small One-loop correction to the ground state energy level which gives
                 the decay rate.}
 \label{fig.5}
\end{figure}
However, the same diagram, but now with a $\pi^0\pi^0$ in the intermediate state, 
is purely imaginary. In the bound state picture it is given by the matrix element
\beq
    \Delta E = {1\over 2} \int\!{d^3p\over (2\pi)^3}\int\!{d^3k\over (2\pi)^3}
   \int\!{d^3q\over (2\pi)^3} \Psi^*(\bp)\, C_0\, 
   {1\over 2\Delta m - \bk^2/m_0 + i\e}\, C_0\, \Psi(\bq)               \label{Delta}
\eeq
as first shown by Labelle\cite{Labelle} using non-relativistic effective field theory.
Here $\Delta m = m_+ - m_-$ gives the energy of the intermediate state and 
\beq
    \Psi(\bp) = {8\pi^{1/2} \gamma^{5/2} \over (\bp^2 + \gamma^2)^2}   \label{Psi}
\eeq
is the Fourier transform of the ground-state wavefunction. It gives the probability to 
find the momentum $p$ in this state. Using the regularized value of the integral 
Eq.(\ref{Ip}), we obtain
\beq
    \Delta E = -i{m_0\over 8\pi}\;C_0^2\,\sqrt{2\Delta m m_0}\, |\Psi(0)|^2
\eeq
This imaginary result signals that the ground state is unstable and will decay with a
rate  $\Gamma = - 2\,\mbox{Im}\,\Delta E$ induced by the hadronic coupling
$C_0$. With the value given in Eq.(\ref{C}) it becomes
\beq
     \Gamma = {16\pi\over 9m_{\pi}^2}|\Psi(0)|^2{m_0^2\over m_{\pi}^2}
              \sqrt{2\Delta m\over m_0} (a_0 - a_2)^2               \label{Gamma}
\eeq
which is the standard result. 

However, there is some implicit uncertainty here in what value to use for the pion 
mass $m_\pi$. It comes from the definition
of the scattering lengths for both charged and neutral pions. Taking it to be
the charged mass $m_+$ for both  of them, we can write the rate as 
$\Gamma = \Gamma_0(1 -3\Delta m/ 2m_+)$ where 
\beq
     \Gamma_0 = {16\pi\over 9m_+^2}|\Psi(0)|^2 \sqrt{2\Delta m\over m_+} (a_0 - a_2)^2
                                                              \label{Gamma0}
\eeq
Since $\Delta m/m_+ = 0.033$, the last factor represents a 5\%  reduction of the 
main decay rate Eq.(\ref{Gamma0}). Such kinematic corrections will be important
in a future experimental determinations of the scattering lengths from the measured
pionium decay rate.

Evaluating the two-loop correction to the ground-state energy, we find a purely real
result since each bubble gives an imaginary contribution. Therefore in the order 
we are working at, the two-loop correction to the decay rate is zero.

The above standard result for the decay rate is due to the constant part of
the $\pi^+ + \pi^- \ra \pi^0 + \pi^0$ amplitude, i.e. the corresponding scattering 
length. But it has also an energy-dependent component parameterized by the
scattering slope $b$ in Eq.(\ref{ReT}). We can now easily calculate this effect
to lowest order in the corresponding derivative coupling $C_2$ in the Lagrangian 
Eq.(\ref{eft}). It results from evaluating the same diagram in Fig.5 but with one 
of the $C_0$ vertices replaced with the $C_2$ vertex from Fig.4. This gives the 
additional contribution
\beq
    \Delta E^{(2)} = - \int\!{d^3p\over (2\pi)^3}\int\!{d^3k\over (2\pi)^3}
\int\!{d^3q\over (2\pi)^3} \Psi^*(\bp)\, C_0\, {1\over 2\Delta m - \bk^2/m_0 + i\e}\,
               {1\over 2}C_2(\bk^2 + \bq^2)\,\Psi(\bq)                 \label{delta}
\eeq
to the ground-state level shift Eq.(\ref{Delta}). The integrations over momenta $\bk$ 
and $\bq$ are now even more divergent than in the first contribution. But again we 
invoke dimensional regularization. We then find that the part coming from the $\bq^2$ 
can be neglected since it is smaller by a factor $\alpha^2 m_0/\Delta m$. Thus the 
integral over $\bq$ just gives $\Psi(0)$. Writing the integral over $\bk$ as
\beq
     \int\!{d^3k\over (2\pi)^3}{\bk^2\over 2\Delta m m_0- \bk^2 + i\e}  
      = \int\!{d^3k\over (2\pi)^3}
     \left[{2\Delta m m_0\over 2\Delta m m_0 - \bk^2 + i\e} - 1\right],    \label{J}   
\eeq
we see that the last term is zero with dimensional regularization while the first part
is just the previous integral Eq.(\ref{Ip}), used in the calculation
of the main level shift. In this way we obtain the finite result
\beq
    \Delta E^{(2)} = i{m_0\over 4\pi}\;C_0 C_2\, 
                \Delta m m_0\sqrt{2\Delta m m_0}\, |\Psi(0)|^2
\eeq
Using now Eq.(\ref{C2}) for the coupling constant $C_2$, we thus obtain
the corresponding correction $\Delta \Gamma^{(2)}$ to the decay rate. It can be written as
\beq
    {\Delta \Gamma^{(2)} \over \Gamma_0} = 2{\Delta m\over m_+}  \left[ {b_0 - b_2\over a_0 - a_2} 
      + (a_0^2 + a_0 a_2 + a_2^2) \right] 
\label{final}
\eeq
when we write $m_0 = m_+$ to leading order in the mass difference. The experimental
values\cite{DGH} of scattering lengths and slopes are $a_0 = 0.26 \pm 0.05$, 
$a_2 = -0.028 \pm 0.012$ and $b_0 = 0.25 \pm 0.03$, $b_2 = -0.082 \pm 0.008$.
Using these, we find that the first term is more than an order of 
magnitude larger than the last. Combined, this amount to a 7.6\% + 0.4\% = 8.0\% correction 
to the main decay rate. On the other hand, using just the tree-level values\cite{DGH},
we obtain the very similar result 8.6\% + 0.1\% = 8.7\%. With values from higher order chiral
perturbation theory, the result is again not much different. The overall hadronic correction
is sizable and larger than other known corrections of electromagnetic origin\cite{many}
\cite{Labelle}.

As a rough check of this rather large correction, we can try to estimate the decay rate
directly from the matrix element Eq.(\ref{ReT}) for $\pi^+ + \pi^- \ra \pi^0 + \pi^0$
with $a = (a_0 - a_2)/3$ and $b = (b_0 - b_2)/3$. Taking $\Gamma \propto |T|^2$, we
obtain to lowest order in the scattering slope the correction factor $1 + 
2{b\over a}{\ex{\bp^2}\over m_\pi^2}$ to the standard result Eq.(\ref{Gamma}). Dividing the average
momentum $\ex{\bp^2}$ equally between the initial state where it is $\gamma^2$ and 
can thus be neglected and the final state where it is $2\Delta mm_\pi$, we have exactly the 
dominant term in the more accurate result Eq.(\ref{final}).

We are grateful to P. Labelle for many discussions on NRQED, pionium and applications 
of effective field theories. Also H. Pilkuhn is acknowledged
for drawing our attention to hadronic atoms. We thank NORDITA for their generous
support and hospitality. Xinwei Kong is supported by the Research Council of Norway.

\end{document}